# Quark-Lepton Complementarity Predictions for $\theta_{23}^{pmns}$ and CP Violation

Gazal Sharma[*] and B. C. Chauhan[†]

*Department of Physics & Astronomical Science,*
*School of Physical & Material Sciences,*
*Central University of Himachal Pradesh (CUHP), Dharamshala, Kangra (HP), India 176215*

**Abstract**

In the light of recent experimental results on $\theta_{13}^{pmns}$, we re-investigate the complementarity between the quark and lepton mixing matrices and obtain predictions for most unsettled neutrino mixing parameters like $\theta_{23}^{pmns}$ and CP violating phase invariants $J$, $S_1$ and $S_2$. This paper is motivated by our previous work where in a QLC model we predicted the value for $\theta_{13}^{pmns} = (9_{-2}^{+1})°$, which was found to be in strong agreement with the experimental results. In the QLC model the non-trivial correlation between CKM and PMNS mixing matrices is given by a correlation matrix ($V_c$). We do numerical simulation and estimate the texture of the $V_c$ and in our findings we get a small deviation from the Tri-Bi-Maximal (TBM) texture and a large from the Bi-Maximal one, which is consistent with the work already reported in literature. In the further investigation we obtain quite constrained limits for $sin^2\theta_{23}^{pmns} = 0.4235_{-0.0043}^{+0.0032}$ that is narrower to the existing ones. We also obtain the constrained limits for the three CP violating phase invariants $J$, $S_1$ and $S_2$: as $J < 0.0315$, $S_1 < 0.12$ and $S_2 < 0.08$, respectively.

---

[*]gazzal.sharma555@gmail.com
[†]chauhan@iucaa.ernet.in

# 1 Introduction

In the past several years the results of neutrino oscillation experiments summarized in particle data group [1] have provided us a very robust evidence that the neutrinos are massive, the lepton flavours are mixed and they oscillate. The most recent result from Daya-Bay and other experiments [2, 3] show relatively large value of $\theta_{13}^{pmns}$, which was also suggested by a number of theoretical and phenomenological analyses [4]. So, the current experimental situation is such that we measured all the quark and charged lepton masses, and the value of the difference between the squares of the neutrino masses $\Delta m_{12}^2 = m_2^2 - m_1^2$ and $|\Delta m_{23}^2| = |m_3^2 - m_2^2|$. We also know the value of the quark mixing angles and the mixing angles $\theta_{12}^{pmns}$, $\theta_{23}^{pmns}$ and $\theta_{13}^{pmns}$ in the lepton sector. What still to be settled are mainly the absolute mass of neutrinos, mass hierarchy in the neutrino mass spectrum (i.e. to determine the sign of $\Delta m_{23}^2$), and nature of neutrinos (Dirac or Majorana), etc.[5]. The precision on $\theta_{23}^{pmns}$ angle, in which quadrant it lies, is another challenge to settle. Finally, as $\theta_{13}^{pmns}$ is non-zero and not too small, there arises a hope to measure the CP violating phases.

In the line of Cabibbo-Kobayashi-Maskawa($U_{ckm}$) mixing matrix in quark sector, the phenomenon of lepton flavour mixing is described by a 3×3 unitary matrix called Pontecorvo-Maki-Nakagawa-Sakata ($U_{pmns}$). Investigating global data fits of the experimental results, so far we have got a picture which suggests that the $U_{pmns}$ matrix contains two large and a small mixing angles; i.e. the $\theta_{23}^{pmns} \approx 45°$, the $\theta_{12}^{pmns} \approx 34°$ and the $\theta_{13}^{pmns} \approx 9°$.

As these results are seen along with the quark flavour mixing matrix ($U_{ckm}$), where all the three mixing angles are small i.e. $\theta_{12}^{ckm} \approx 13°$, $\theta_{23}^{ckm} \approx 2.4°$ and $\theta_{13}^{ckm} \approx 0.2°$, a disparity-cum-complementarity between quark and lepton mixing angles is noticed. Since, the quarks and leptons are fundamental constituents of matter and also that of particles' Standard Model(SM), the complementarity between the two families is seen as a consequence of a symmetry at some high energy scale. This complementarity popularly named 'Quark-Lepton Complementarity'(QLC) has been explored by several authors [6]-[11]. The relation is quite appealing to do the theory and phenomenology, however it is still an open question, what kind of symmetry could be there between these fundamental particles of two sectors.The possible consequences of QLC have been widely investigated in the literature. In particular a simple correspondence between the $U_{pmns}$ and $U_{ckm}$ matrices has been proposed and used by several authors [12]-[15] and analysed in terms of a correlation matrix [16]-[22].

The main motivation of the present work is to re-visit the QLC Model, which we proposed in 2007 [23]. The work came into light, when its predictions for the reactor mixing angle $\theta_{13}^{pmns} = (9^{+1}_{-2})°$ was found to be in strong agreement with the experimental results [2, 3]. In the light of recent reactor angle data and updated experimental statistics on the other mixing parameters it was necessary to work again on the model, investigate and update the predictions. Keeping in view the fact that the neutrino physics has entered in the precision era, we chose a more accurate Wolfenstein parametrization for $U_{ckm}$ i.e. preserving unitarity upto the sixth order -$\mathcal{O}(\lambda^6)$- so called Next-to-Leading order [24].

Using all these ingredients we re-investigate the probability density textures of the correlation matrix $V_c$, numerically. We found that the most favoured pattern of $V_c$ slightly deviates from TriBi-Maximal (TBM) and largely from Bi-Maximal (BM) one. As per our investigations a clear non-trivial structure of $V_c$ and the strong indication of gauge coupling unification at high scale helped us to put constraints on the most un-settled angle $\theta_{23}^{pmns}$ and on unknown CP violation invariants $J$, $S_1$ and $S_2$.

In the next section (**2**), we describe in brief the theory of the QLC model and the investigation of correlation matrix ($V_c$) using Monte Carlo simulation is done in section (**3**). The verification of our previous work with $(V_c)^{13} = 0$ are checked and an updated version of $V_c$ matrix texture is obtained in the same section. As per the model procedure, using the most probable texture of the correlation matrix we derive the constraints on the $\theta_{23}^{pmns}$ mixing angle and lepton CP violating phase invariants $J$, $S_1$, and $S_2$ in the section (**4**). Finally, the conclusions are summarized in the section (**5**).

## 2  Theory of the QLC Model

The Standard Model of particle physics (SM) provides the best theoretical description of physical world at energies so far probed by experiments. The neutrinos are the only fermions in SM without right-handed partners and any electrical charge. Since they do not have any right handed partners, they are massless. However, the recent experimental results show that they mix and do oscillate, such that they are massive and have relatively small masses. This suggests a need to learn physics beyond the SM framework i.e. any new phenomena must appear at some scale associated with the existence of neutrino masses and mixing. Any realistic model, such as see-saw model, must produce such tiny non-zero masses and large mixing when reduced to an effective low energy theory. Thus it is quite possible that SM of particle interactions is a low-energy limit of some underlying theory whose true structure will emerge only when higher energy scales are probed. Certainly a deep understanding of the algebraic relationship between quarks and leptons at high energy scale will be interesting. At the same time the stability check of the RGE effects on the model equation is also essential.

The flavour mixing stems from the mismatch between the left handed rotations of the up-type and down-type quarks, and the charged leptons and neutrinos. Such mixing of quarks and leptons has always been of great interest and remains a mystery in particle physics. In the SM the mixing of quark and lepton sectors is described by the matrices $U_{ckm}$ and $U_{pmns}$, respectively, which show up in the charged current interaction described by the Lagrangian as:

$$L = -\frac{g}{\sqrt{2}} q1_L^\dagger \gamma^\mu U_{\text{ckm}} q2_L W_\mu^+ - \frac{g}{\sqrt{2}} l1_L^\dagger \gamma^\mu U_{\text{pmns}} l2_L W_\mu^- + h.c. \tag{1}$$

where

$$q1_L = -(u_L, c_L, t_L)^T; q2_L = (d_L, s_L, b_L)^T; l1_L = -(e_L, \mu_L, \tau_L)^T; l2_L = (\nu_1, \nu_2, \nu_3)^T. \tag{2}$$

This relation when viewed with the observed pattern of mixing angles of quarks and leptons and combined with the pursuit for unification i.e. symmetry at some high energy leads the concept of quark lepton complementarity i.e. QLC. To look for such unification, it is useful to work in a basis where the quark and lepton Yukawa matrices are related. In general, the Yukawa matrices for quarks are taken as $Y_u$ and $Y_d$ i.e. for up and down quark sectors, respectively. The diagonalizing matrices are given by

$$Y_u = U_u Y_u^\Delta V_u^\dagger \quad \text{and} \quad Y_d = U_d Y_d^\Delta V_d^\dagger, \tag{3}$$

where the $Y^\Delta$ are diagonal and the $U$ and $V$ are unitary matrices. Such that the observable quark mixing matrix $U_{ckm}$ is given by

$$U_{ckm} = U_u^\dagger U_d. \tag{4}$$

However, for the charged lepton sector, the Yukawa matrix is given by

$$Y_l = U_l Y_l^\Delta V_l^\dagger. \tag{5}$$

For neutrino sector, we introduce one right-handed singlet neutrino per family; i.e. $M_R$ as the Majorana mass matrix for the right handed neutrino and $M_D$ the Dirac mass matrix. This leads to light neutrino masses given by the Type-I see-saw mechanism after breakdown of electroweak symmetry

$$M_\nu = M_D \frac{1}{M_R} M_D^T = (U_0 M_D^\Delta V_0^\dagger) \frac{1}{M_R} (V_0^\star M_D^\Delta U_0^T), \tag{6}$$

where $U_l$, $V_l$ and $U_0$, $V_0$ diagonalize the charged lepton and $M_D$, respectively.

The neutrino mass matrix can be rewritten as

$$M_\nu = U_0 V_c M_\nu^\Delta V_c^T U_0^T, \tag{7}$$

where $V_c$ represents the rotation of $M_D^\Delta V_0^\dagger \frac{1}{M_R} V_0^\star M_D^{\Delta T}$.

The mixing matrix $V_c$ is here defined to verify the equality $U_\nu \equiv U_0 V_c$ and is such that in lines of $V_{ckm}$ the lepton mixing matrix is given by

$$U_{pmns} = U_l^\dagger U_\nu = U_l^\dagger U_0 V_c \,. \tag{8}$$

In grand unification, the down-type quarks and the charged leptons are in general assigned into a multiplet, we assume that the following simple relations hold

$$Y_l \approx Y_d^T \to U_l \simeq V_d^\star \,. \tag{9}$$

In the same way, if we call $Y_\nu$ the Yukawa coupling that will generate the Dirac neutrino mass matrix $M_D$, we have also the relation

$$Y_\nu \approx Y_u^T \quad \to \quad U_0 \simeq V_u^\star \,. \tag{10}$$

Such that

$$U_{pmns} = V_d^T U_0 V_c = V_d^T V_u^* V_c \,. \tag{11}$$

If we further assume $Y_\nu \approx Y_u$, which can be realized in some larger gauge group such as generic SO(10). In addition, for symmetric form of the down-type quark Yukawa matrix, there arises an interesting relationship amongst the quark mixing, lepton mixing and the correlation matrices $V_c$. Combining the relations derived for quark and lepton we get

$$Y_u \approx Y_u^T \quad \to \quad U_u \simeq V_u^\star \,. \tag{12}$$

and ,Using the equations we get

$$U_{pmns} \simeq U_{ckm}^\dagger V_c \,. \tag{13}$$

The form of $V_c$ can be obtained under some assumptions about the flavor structure of the theory

$$V_c = U_{ckm} \cdot \Psi \cdot U_{pmns} \tag{14}$$

where the quantity $\Psi$ is a diagonal matrix $\Psi = \text{diag}(e^{\iota \psi_i})$ and the three phases $\psi_i$ are free parameters as they are not restricted by present experimental evidences. This is more appropriate to do because in Grand Unified Theories (GUTs), once quarks and leptons are inserted in the same representation of the underlying gauge group, in order to counter the phase mismatch one has to include arbitrary but non-trivial phases between the quark and lepton mixing matrices.

The simplest possiblity for the correlation matrix, which has been widely explored in literature is $V_c = U_{ckm} \cdot U_{pmns}$. Here the correlation matrix $V_c$ is taken bi-maximal in nature such that two angles of 45° and one 0°. Such that, the original form proposed for QLC relation was

$$\theta_{12}^l + \theta_{12}^q = 45°, \quad \theta_{23}^l + \theta_{23}^q = 45°, \quad \theta_{13}^l + \theta_{13}^q = 0°, \tag{15}$$

where $\theta_{ij}^{l,q}$ are the respective lepton and quark mixing angles.

However, in our model we adopt a more general and democratic approach for the correlation matrix i.e. it may take any form of texture as suggested by the input form theory and experimental data from quark and lepton sectors.

## 3 Numerical Analysis

In this section we investigate the texture of $V_c$ correlation matrix taking into account the experimental updates and the Wolfenstein parametrization for $U_{ckm}$ i.e. unitary up to $\mathcal{O}(\lambda^6)$ also called as Next-to-Leading order [24] as

$$U_{ckm} = \begin{bmatrix} 1 - \lambda^2/2 - \lambda^4/8 & \lambda & A\lambda^3(1 + \lambda^2/2)(\bar{\rho} - \iota\bar{\eta}) \\ -\lambda + A^2\lambda^5(1/2 - \bar{\rho} - \iota\bar{\eta}) & 1 - \lambda^2/2 - \lambda^4/8(1 + 4A^2) & A\lambda^2 \\ A\lambda^3(1 - \bar{\rho} - \iota\bar{\eta}) & -A\lambda^2 + A\lambda^4(1/2 - \bar{\rho} - \iota\bar{\eta}) & 1 - A^2\lambda^4/2 \end{bmatrix} + \mathcal{O}(\lambda^6).$$

The Wolfenstein parameters $\lambda, A, \rho, \eta$ are

$$\sin\theta_{12}^{ckm} = \lambda \tag{16}$$

$$\sin\theta_{23}^{ckm} = A\lambda^2 \tag{17}$$

$$\sin\theta_{13}^{ckm} e^{-\delta^{ckm}} = A\lambda^3(\rho - \iota\eta) \tag{18}$$

where

$$\rho + \iota\eta = \frac{\sqrt{1 - A^2\lambda^4}(\bar{\rho} + \iota\bar{\eta})}{\sqrt{1 - \lambda^2}[1 - A^2\lambda^4(\bar{\rho} + \iota\bar{\eta})]} \tag{19}$$

On the other hand the lepton mixing matrix $U_{pmns}$ is parametrized as [23]

$$U_{pmns} = U_{23}.\phi.U_{13}.\phi^\dagger.U_{12}.\phi_m. \tag{20}$$

Here $\phi \sim diag(1, 1, e^{\iota\phi})$ and $\phi_m \sim diag(e^{\iota\phi_1}, e^{\iota\phi_2}, 1)$ are diagonal matrices containing the Dirac and Majorana CP violating phases, respectively. Such that

$$U_{pmns} = \begin{bmatrix} e^{\iota\phi_1}c_{12}c_{13} & e^{\iota\phi_2}c_{13}s_{12} & s_{13}e^{-\iota\phi} \\ e^{\iota\phi_1}(-c_{23}s_{12} - e^{\iota\phi}c_{12}s_{13}s_{23}) & e^{\iota\phi_2}(c_{12}c_{23} - e^{\iota\phi}s_{12}s_{13}s_{23}) & c_{13}s_{23} \\ e^{\iota\phi_1}(-e^{\iota\phi_1}c_{12}c_{23}c_{13} + s_{12}s_{23}) & e^{\iota\phi_2}(-e^{\iota\phi}c_{23}s_{12}s_{13} - c_{12}s_{23}) & c_{13}c_{23} \end{bmatrix}$$

.

However, the values of the $U_{ckm}$ parameters [1] and $U_{pmns}$ angles[25] are taken as under at 1-$\sigma$ level

$$\lambda = 0.2255 \pm 0.0006, \quad (21)$$
$$A = 0.818 \pm 0.015,$$
$$\bar{\rho} = 0.124 \pm 0.024,$$
$$\bar{\eta} = 0.354 \pm 0.015$$

$$\sin^2\theta_{13} = 0.0218^{+0.0010}_{-0.0010} \quad (22)$$
$$\sin^2\theta_{12} = 0.304^{+0.013}_{-0.012}$$
$$\sin^2\theta_{23} = 0.452^{+0.052}_{-0.028}$$
$$\phi = (306°)^{+39}_{-70}.$$

For the unknown phases $\phi_1$, $\phi_2$ and the three $\psi_i$ (**14**), as they are not constrained by any experimental data, we vary their values between the interval $[0, 2\pi]$ in a flat distribution.

We perform the investigation by using the numerical method of Monte Carlo simulation, which generated one billion i.e. $10^8$ values for each variable with two-sided Gaussian distributions around the mean values of the observables including the quark parameters $A$, $\lambda$, $\bar{\rho}$, $\bar{\eta}$. Such that we obtained fine, smooth histogram for probability density of each elements of the $V_c$.

In our previous work we have seen that the 13-element of the correlation matrix is strongly weighted to zero, namely $V_c^{13} = 0$. Such that the possibility for $V_c$ to be BM, TBM or any other with 13-element quite small was quite open. However, with the inclusion of recent data for $\theta_{13}^{pmns}$ and other mixing parameters in the present work we find a little bit different $V_c$ i.e. a small deviation form the TBM pattern and large deviation from the BM texture. This can be seen by the matrices and a figure of histogram panels given below:

$$V_c = \begin{bmatrix} 0.68...0.92 & 0.37...0.68 & 0.0004...0.30 \\ 0.14...0.68 & 0.36...0.78 & 0.56...0.74 \\ 0.24...0.46 & 0.48...0.67 & 0.66...0.77 \end{bmatrix}$$

The corresponding histograms of probability density distribution for all the 9-elements of $V_c$ matrix are shown in figure **1** as the respective nine panels.

In these histograms we have compared the $V_c$ matrix with the BM and TBM structure of matrices. The solid lines correspond to the TBM matrix and the dashed line in all the panels represents the corresponding BM matrix element. It can be seen from the figure(s) that the $V_c$ generated is deviated slightly from TBM and largely from BM. This has also been reported earlier in the literature [26]. There are panels in the figure corresponding to 13, 23, 33 elements in which the dashed and solid lines are overlapped, as their values are exactly same.

On comparing the weighted value of each element taken from the histograms, in figure **1**, of $V_c$ matrix with corresponding elements of BM and TBM matrices, we can see as

$$V_c = \begin{bmatrix} 0.80 & 0.54 & 0.18 \\ 0.44 & 0.58 & 0.66 \\ 0.36 & 0.58 & 0.72 \end{bmatrix} \qquad BM = \begin{bmatrix} 0.71 & 0.71 & 0.00 \\ 0.50 & 0.50 & 0.71 \\ 0.50 & 0.50 & 0.71 \end{bmatrix}$$

and

$$V_c = \begin{bmatrix} 0.80 & 0.54 & 0.18 \\ 0.44 & 0.58 & 0.66 \\ 0.36 & 0.58 & 0.72 \end{bmatrix} \qquad TBM = \begin{bmatrix} 0.82 & 0.58 & 0.00 \\ 0.41 & 0.58 & 0.71 \\ 0.41 & 0.58 & 0.71 \end{bmatrix}$$

In the matrices and the 9-panels of the figure **1** we can clearly see that $V_c$ matrix deviates from BM structure in almost all the elements except 21, 22 and 33. It deviates largely from 11, 12, 13, 23, 31, and 32 elements. On the other hand, for TBM pattern it deviates significantly only for 13, 23 and 31 elements. So we can make a statement that there is large deviation from BM matrix and a slight deviation from the TBM texture. It may be noted that the significant deviation in the 13 element of $V_c$ matrix i.e. 0 to 0.18, such that $V_c \neq 0$ does not contradict our previous work rather generalize it further. To check and verify this analytically, numerically and graphically we obtain the following expression relating the PMNS angle $\theta_{13}^{pmns}$ (LHS) in terms of the angles of the

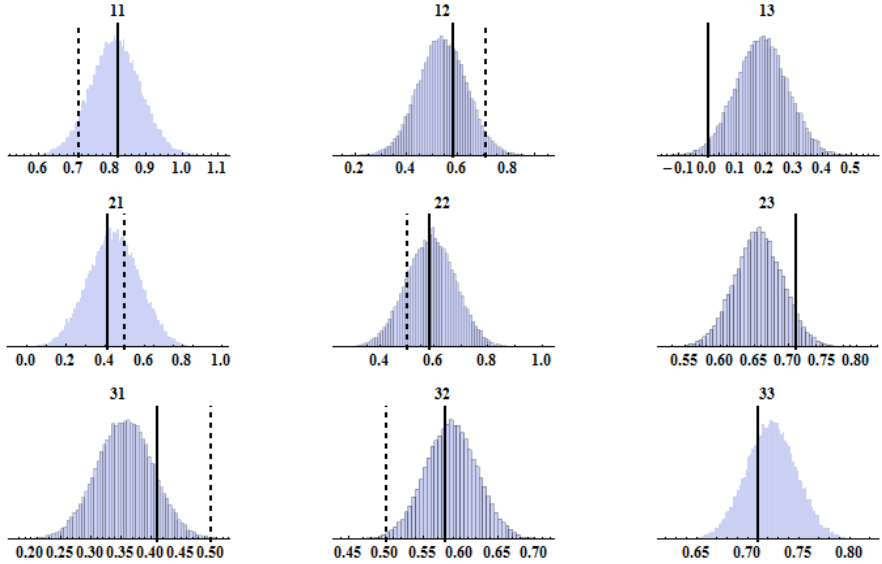

*Figure 1: Probability density distribution of all the 9 elements of $V_c$ matrix superimposed by the corresponding elements of BM (Dashed) and TBM (solid) lines*

correlation matrix $V_c$ (RHS):

$$\begin{aligned}
\sin^2 \theta_{13}^{pmns} &= \sin^2 \theta_{13} - 2(e^{(-\iota\phi)} \cos\theta_{13} \sin\theta_{13} \sin\theta_{23})\lambda \\
&+ (-\sin^2 \theta_{13} + \cos^2 \theta_{13} \sin^2 \theta_{23})\lambda^2 \\
&+ e^{(-\iota\phi)} \cos\theta_{13} \sin\theta_{13}(2A\cos\theta_{23} + 2\iota A \cos\theta_{23}\bar{\eta} \\
&- 2A\cos\theta_{23}\bar{\rho} + \sin\theta_{23})\lambda^3 + 2A\cos^2\theta_{13}\cos\theta_{23} \\
&(-1 - \iota\bar{\eta} + \bar{\rho})\sin\theta_{23}\lambda^4 + 1/4 e^{(-\iota\phi)} \cos\theta_{13} \sin\theta_{13} \\
&(-4A\cos\theta_{23} - 4\iota A\cos\theta_{23}\bar{\eta} + 4A\cos\theta_{23}\bar{\rho} + \sin\theta_{23} + 4A^2 \sin\theta_{23} \\
&+ 8\iota A^2 \bar{\eta}\sin\theta_{23} - 8A^2 \bar{\rho}\sin\theta_{23})\lambda^5 + \mathcal{O}(\lambda^6)
\end{aligned} \quad (23)$$

We test our previous results by taking $\sin^2 \theta_{13} = 0$ for $V_c$ and CKM matrix up to the order $\mathcal{O}(\lambda^3)$, such that the above equation for $\sin^2 \theta_{13}^{pmns}$ becomes

$$\sin^2 \theta_{13}^{pmns} = \sin^2 \theta_{23} \lambda^2 + \mathcal{O}(\lambda^3)$$

.

We can see that this is exactly the equation **21** of the work [23]. Furthermore for the CKM matrix up to the order $\mathcal{O}(\lambda^6)$ the above equation becomes

$$\sin^2 \theta_{13}^{pmns} = \sin^2 \theta_{23}\lambda^2 + 2A\cos\theta_{23}(-1 - \iota\bar{\eta} + \bar{\rho})\sin\theta_{23}\lambda^4 + \mathcal{O}(\lambda^6). \quad (24)$$

From the figure we can clearly see that the allowed range for $\sin^2 \theta_{13}$ of $V_c$ corresponding to the experimental range of $\sin^2 \theta_{13}^{pmns}$ is quite wide including $\sin^2 \theta_{13} = 0$. This gives a complete justification for the result $V_c^{13} = 0$ which is also consistent with the experimental value ($\approx 9°$) and as we reported in our previous work.

The graphical representation of the most general expression is shown in the figure **2**. The vertical lines superimposed over are the 1-$\sigma$ range of the recent experimental data for $\sin^2 \theta_{13}^{pmns}$ from [2, 3].

## 4  Results and Discussions

Unlike our previous work, we see a probabilistic departure from the BM and TBM pattern, i.e. specifically $V_c^{13} \neq 0$. The reason behind such a situation is that the estimation of this element $V_c^{13}$ is weighted heavily by the value $\theta_{13}^{pmns}$ as compared to the other PMNS angles and CKM parameters. So, keeping $\theta_{13}^{pmns}$ free to vary flatly between 0.00 to 0.32 as we did in [23] the non-trivial combination PMNS angles, $\lambda$'s and other non-restricted phases etc. leads to the probability density of the element $V_c^{13}$ to peak at zero. However, for a very restricted Gaussian range for $\theta_{13}^{pmns}$ the peak of probability distribution of the element $V_c^{13}$ gets shifted to 0.18, as in the present study.

It may be noted that there exists some flavor models, which imply a correlation $V_c$ with $(V_c)^{13} = 0$. In some discrete flavor symmetries such as $A_4$ dynamically broken into $Z_3$, as in [27] and [28], or $S_3$ softly broken into $S_2$, as in [29], the TBM structure appears naturally. However, in the present case where $V_c$ deviates from BM and TBM structure i.e. $V_c^{13} \neq 0$, we need to look for some relevant flavour models.

As per our model procedure, in order to constrain the lepton mixing parameter namely the less constrained ones $\theta_{23}^{pmns}$ and the three lepton CP violating phase invariants ($J$, $S_1$, $S_2$) we use the inverse equation [23]

$$U_{pmns} = (U_{ckm} \cdot \Psi)^{-1} \cdot V_c. \tag{25}$$

This expression is the inverse of equation **14**, which was used to estimate the texture of the correlation matrix $V_c$. We used the weighted values of the matrix elements in equation **25** and full spread $[0, 2\pi]$ of the unconstrained phase angles $\phi$ & $\psi$, we resorted to Monte Carlo simulation, which genrated one billion values for each variable.

## 4.1 Predictions for $\theta_{23}^{pmns}$

In this section we investigate the implications of the non-trivial structure of the $V_c$ correlation matrix in the light of the latest results of $\theta_{13}^{pmns}$.

After using parametrization the $V_c$ from equation **14**, we can find analytical equation, which connects $V_c$ angles on RHS with $\sin^2 \theta_{23}^{U_{pmns}}$ on LHS. So, from equation **25** we derive expression for $U_{23}^{pmns}$ that includes $\lambda$-terms upto $6^{th}$ order on the RHS (all the angles are pertaining to $V_c$) is

$$\begin{aligned}
U_{23}^{pmns} &= 2e^{(-\iota\phi-2\iota\psi 2)}\lambda\cos\theta_{13}\sin\theta_{13}\sin\theta_{23} + e^{(-2\iota\psi 2)}\cos^2\theta_{13}\sin^2\theta_{23} - e^{(-\iota\phi-2\iota\psi 2)}\lambda^3 \\
&\quad \cos\theta_{13}\sin\theta_{13}(2A\cos\theta_{23} + \sin\theta_{23}) - 1/4 e^{(-\iota\phi-2\iota\psi 2)}\lambda^5\cos\theta_{13}\sin\theta_{13} \\
&\quad (-4A\cos\theta_{23} - 8\iota A\cos\theta_{23}\bar{\eta} + 8A\cos\theta_{23}\bar{\rho} + \sin\theta_{23} \\
&\quad +4A^2\sin\theta_{23}) + Ae^{(-2\iota\psi 2)}\lambda^4\cos^2\theta_{13}(A\cos^2\theta_{23} \\
&\quad +2\cos\theta_{23}\sin\theta_{23} + 2\iota\cos\theta_{23}\bar{\eta}\sin\theta_{23} - 2\cos\theta_{23}\bar{\rho}\sin\theta_{23} - A\sin^2\theta_{23}) \\
&\quad +1/256 e^{(-2\iota(\phi+\psi 2))}\lambda^2(256\sin^2\theta_{13} - 256 e^{(2\iota\phi)} \\
&\quad \cos^2\theta_{13}\sin\theta_{23}(2A\cos\theta_{23} + \sin\theta_{23})). \quad (26)
\end{aligned}$$

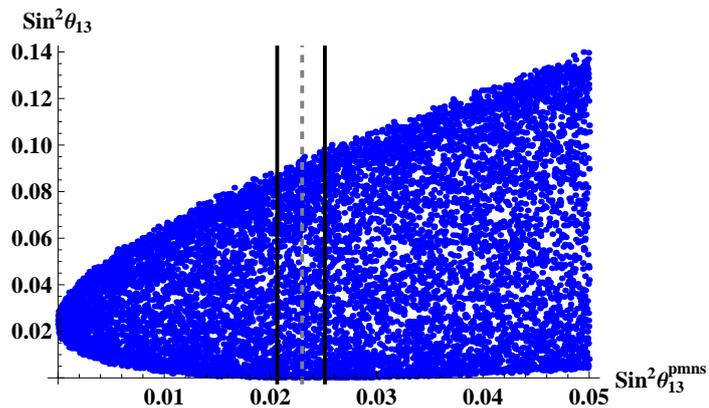

Figure 2: The figure shows the allowed parameter space between $\sin^2 \theta_{13}$ of $V_c$ and $\sin^2 \theta_{13}^{pmns}$

After simplifying equation **26** we can obtain equation for $\sin^2\theta_{23}^{pmns}$ upto $4^{th}$ order i.e. so-called Leading order as

$$\begin{aligned}
\sin^2\theta_{23}^{pmns} &= \sin^2\theta_{23} + 2e^{(-\iota\phi)}\sin\theta_{23}\tan\theta_{13}\lambda \\
&\quad + \sin\theta_{23}(2A\cos\theta_{23} + \sin\theta_{23}) + \tan^2\theta_{13})\lambda^2 \\
&\quad - e^{(-\iota\phi)}(2A\cos\theta_{23} + \sin\theta_{23})\tan\theta_{13}\lambda^3 + \mathcal{O}(\lambda^4). \quad (27)
\end{aligned}$$

The same equation for the Next-to-Leading order i.e. upto $\mathcal{O}(\lambda^6)$ is written as

$$\begin{aligned}
\sin^2\theta_{23}^{pmns} &= \sin^2\theta_{23} + 2e^{(-\iota\phi)}\sin\theta_{23}\tan\theta_{13}\lambda \\
&\quad + \sin\theta_{23}(2A\cos\theta_{23} + \sin\theta_{23}) + \tan^2\theta_{13})\lambda^2 \\
&\quad - e^{(-\iota\phi)}(2A\cos\theta_{23} + \sin\theta_{23})\tan\theta_{13}\lambda^3 + A(A\cos 2\theta_{23} \\
&\quad +2\cos\theta_{23}\sin\theta_{23} + \iota\bar{\eta}\sin 2\theta_{23} - \bar{\rho}\sin 2\theta_{23})\lambda^4 - 1/4(e^{(-\iota\phi)}(-4A\cos\theta_{23} \\
&\quad -8\iota A\cos\theta_{23}\bar{\eta} + 8A\cos\theta_{23}\bar{\rho} + \sin\theta_{23} + 4A^2\sin\theta_{23})\tan\theta_{13})\lambda^5 + \mathcal{O}(\lambda^6). \quad (28)
\end{aligned}$$

From the above equation we obtain histograms of the probability density functions for $\sin^2\theta_{23}^{pmns}$ upto $4^{th}$ order and $6^{th}$ order approximations and show comparison between the two. In the figure **3** we show this quite nicely- the left panel of the figure is for $\sin^2\theta_{23}^{pmns}$ upto $6^{th}$ order (darker shaded), where as the right panel depicts superimposition of the $6^{th}$ order (quite constrained) histogram on the $4^{th}$ order approximation (light shaded).

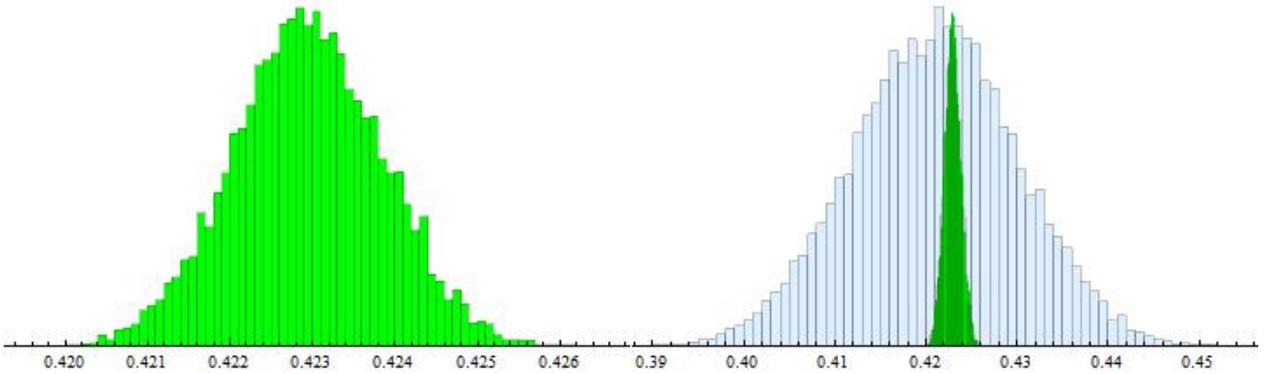

*Figure 3: Probability density distribution of $\sin^2\theta_{23}^{pmns}$ for Next-to-Leading order (left) & its comparison with the same for Leading order*

We measure the improvement in precision of the predictions of angle and it is remarkable, i.e. as we go from $4^{th}$ order to the $6^{th}$ order the 1-$\sigma$ region of the angle is shrank by about 90%. The quite constrained value for $\sin^2\theta_{23}^{pmns}$ pertaining to the Next-to-Leading order is obtained as

$$\sin^2\theta_{23}^{pmns} = 0.4235^{+0.0032}_{-0.0043}$$

We have also verified the accuracy of these numerical results independently solving analytical equation **28**. The values so obtained are as

$$\sin^2\theta_{23}^{pmns} = 0.4054^{+0.0291}_{-0.0572},$$

which lies within $1\sigma$ value.

In the figure **4** we compare our results with the values of Global data analysis given by the various groups [4]. This figure can be further compared with results of [21] where the similar comparison is done for QLC Leading and Next-to-Leading order. As such, two things seen are quite common to discuss that by considering the higher order approximation in the CKM matrix the 1-$\sigma$ ranges of $\sin^2\theta_{23}^{pmns}$ values is constricted significantly and are consistent with all the data fits available, but quite constrained.

### 4.2 CP violating invariants in the lepton sector

We also investigate the consequences of the non-trivial structure of the $V_c$ correlation matrix upon the undetermined CP invariants in the lepton sector. As we know that there are two kinds of CP invariants parametrizing CP violating effect in the leptonic sector. Analogous to the quark sector Jarlskog invariant $J$ that parametrizes the effects related to the Dirac phase, and two additional CP invariants $S_1$ and $S_2$ that parametrize the effects related to the Majorana phases, which arise if neutrinos are Majorana particles. The $J$ invariant describes all CP breaking observables in neutrino oscillations. The most general form of this is given as

$$J = Im\{U_{\nu_e\nu_1}U_{\nu_\mu\nu_2}U^*_{\nu_e\nu_2}U^*_{\nu_\mu\nu_1}\}. \tag{29}$$

Using the parametrization **20** the equation becomes

$$J = \frac{1}{8}\sin 2\theta_{12}\sin 2\theta_{13}\sin 2\theta_{23}\cos\theta_{13}\sin\phi; \tag{30}$$

Similarly, the other two CP invariants $S_1$ and $S_2$ are related to Majorana phases, as under

$$S_1 = \frac{1}{2}\cos\theta_{12}\sin 2\theta_{13}\sin(\phi_1+\phi) \tag{31}$$

$$S_2 = \frac{1}{2}\sin\theta_{12}\sin 2\theta_{13}\sin(\phi_2+\phi). \tag{32}$$

Here the mixing angles and phases involved on the RHS of the CP invariants are obviously corresponding to the PMNS mixing matrix. From the above equations we can see that the two Majorana phases appear in $S_1$ and $S_2$ but not in J. Using simulation for all the three equations of CP invariants, the effect of non-zero but sizeable value of $\theta_{13}$ and updated values of $\theta_{12}$ can be seen in the correlation plots using our model values. The plots shown in the corresponding figures **5,6** are depicting the correlation between Jarlskog invariant $J$ with $\sin^2\theta_{23}$ (upper panel) and $\sin^2\theta_{12}$ (lower panel), and other two invariants $S_1$ (left panel) and $S_2$ (right panel) with $\sin^2\theta_{12}$.

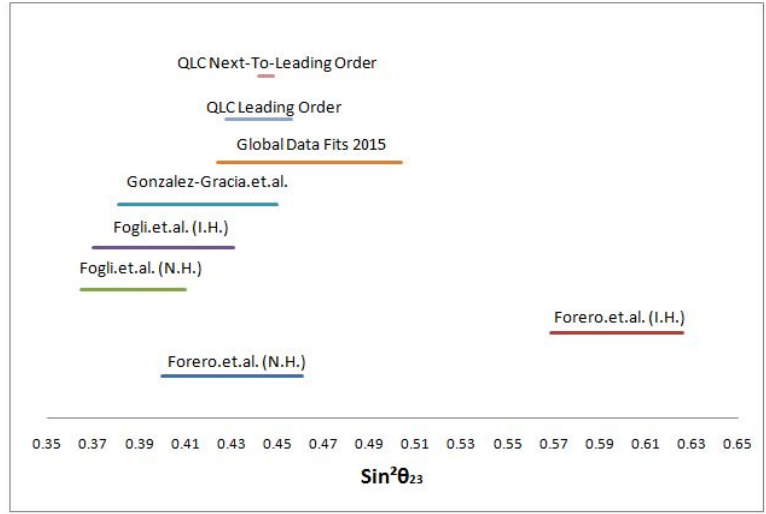

*Figure 4: The 1-$\sigma$ ranges of $\sin^2\theta_{23}^{pmns}$ given by a number of Global data fits plotted with the model value ranges obtained for Leading and Next-to-Leading orders*

From figures we can conclude that the absolute values of the CP violating invariants (J, S$_1$, S$_2$) are constrained as under

|J |< 0.0315;    |S$_1$|< 0.12;    |S$_2$|< 0.08.

The above results can compared to our previous work [22] where J assumes any value between -0.04 & +0.04 and $|S_1| < 0.14, |S_2| < 0.11$ at $1-\sigma$ level for BM and TBM structure of the $V_c$.

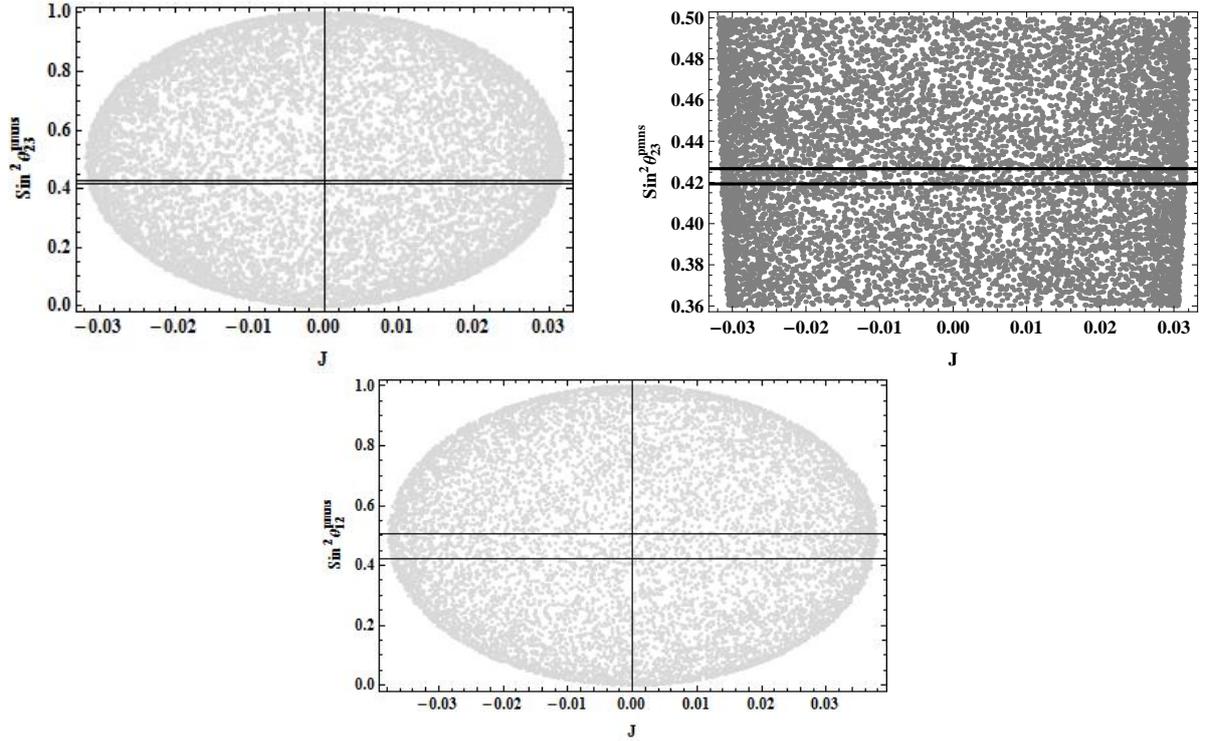

Figure 5: *Correlation between J v/s* $\sin^2\theta_{23}$ *and* $\sin^2\theta_{12}$ *for full range.*

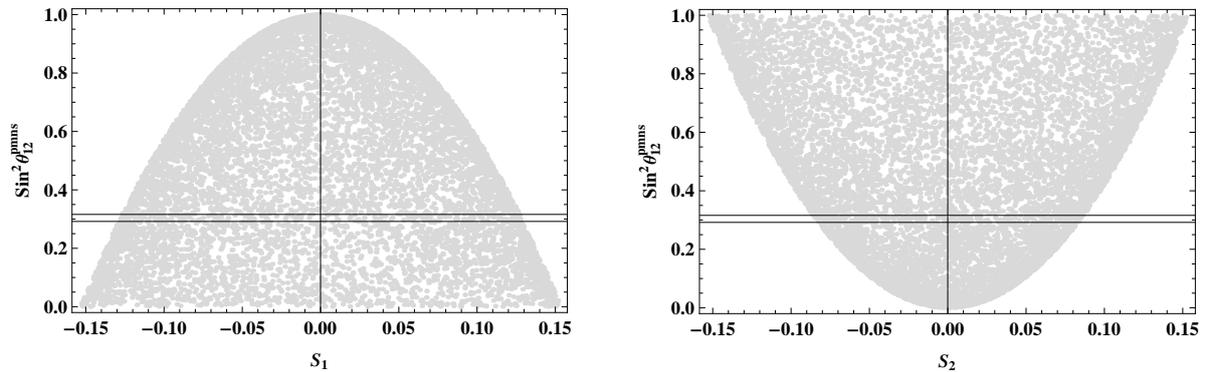

Figure 6: *Correlation between* $S_1$ *and* $S_2$ *v/s* $\sin^2\theta_{12}$ *for full range.*

Thus it is clear that our present ranges for the CP invariants are narrower i.e. $J$ by 21%, $S_1$ by 14% and $S_2$ by 27%.

# 5  Conclusions

The recent non-zero value ( ($9°$)) of reactor mixing angle ($\theta_{13}^{pmns}$) from Daya Bay, RENO and other experiments [2, 3] was found to be in strong agreement with the one predicted by our group [23] in 2007. This drew quite a large attention of scientific community towards the model we used. We took benefit of recent experimental developments and re-investigated our model i.e. mainly the correlation between the $U_{ckm}$ quark and $U_{pmns}$ lepton mixing matrices. In fact, the advantages of the present work over the similar other ones existing in literature e.g. the most recent one of Junpei in [24] are quite remarkable. In fact, we had the following motivation points to continue: 1) Model prediction already tested by the experiments; 2) Experimental updates and global data fits available, specifically on the value of $\theta_{13}^{pmns}$; 3) Wolfenstein parametrization up to the order $\mathcal{O}(\lambda^6)$ to obtain better precision; 4) Unlike other work it was a numerical study supplemented by the analytical one; 5) Quite general and natural in procedure.

As such, it was imperative to re-visit the model, update the things and get some more insight for further predictions. Being most general and natural in procedure, i.e. not restricting the phase mismatch matrix, the disadvantage of this model is that it gives too broad range of parameters even if the experimental data in hand have had quite a large precision. Despite of the fact, after performing rigorous numerical computations and analytical study, we have got interesting results to share with the scientific community.

In a detailed analysis we estimated and found the texture of the correlation matrix $V_c$, which is slightly deviated from a TriBi-Maximal pattern, and largely with a Bi-Maximal pattern. This observation that $V_c$ is much closer to TBM than BM has been also reported in literature [26]. This conclusion endorsed the results of the previous studies [16] and is in agreement with other qualitative arguments that favor the CKM matrix to measure the deviation of the PMNS matrix from exact Bimaximal mixing [30].

The most interesting result and conclusion we could draw from this work is the quadrant of $\theta_{23}^{pmns}$, which is significantly shifted to $< 45°$, i.e. about 5 $\sigma$ below the maximal mixing.

$$sin^2\theta_{23}^{pmns} = 0.4235^{+0.0032}_{-0.0043}. \tag{33}$$

$$\theta_{23}^{pmns} = 40.60^{+0.1°}_{-0.3°}. \tag{34}$$

We obtained a notable improvement in precision of the predictions of angle as as we moved from $4^{th}$ order to the $6^{th}$ order approximation i.e. the 1-$\sigma$ region of $\theta_{23}^{pmns}$ is constrained by about 90%.

Another consequences of the model are the predictions for CP violating phase invariants $J$, $S_1$ and $S_2$; i.e.

$$|J| < 0.0315 \qquad |S_1| < 0.12 \qquad |S_2| < 0.08,$$

which are constrained by 21%, 14%, and 27%, respectively as compared to the previous results.

The observed deviation from TBM and BM values can lead to violation of the TBM conditions and that is to deviating significantly from the TMB form of neutrino mass matrix. It can be a manifestation of some other symmetry existing at high energy scale. In other words, we obtain a texture, which suggests a new kind of flavour symmetry at high scale or TBM broken at low energy scale or could be a weakly broken TBM at the high scale. This opens up several new question to be addressed.

As far as the stability of the QLC relation is concerned, the kind of model relation obtained from the GUT models with some flavor symmetry has been checked by the several works and has been concluded that they are stable under the RGE effects and the radiative corrections are small for supersymmetric parameter $\tan\beta \leq 40$ [31]-[34]. It is quite possible that the QLC relation holds good at the unification scale after counting RGE effects such as for $\tan\beta \leq 40$ the deviation in the $V_c$ from TBM structure is introduced. If supersymmetry is discovered with $\tan\beta \leq 40$ and that would be a strong hint for some relevant flavor symmetry models and their specific Higgs pattern. The future test of the model will be the results from neutrino experiments, in particular, regarding determination of the quadrant of $\theta_{23}$ and the observance of CP violation, and relevant flavour symmetry found at the high energy scale.

# Acknowledgments

We thank Lal Singh for his useful computational help in the beginning of the work. A lot of thanks to IUCAA for providing research facilities during the completion of most part of this work.